# Visualizing Van der Waals Epitaxial Growth of Two-Dimensional Heterostructures

*Kenan Zhang, Changchun Ding, Baojun Pan, Zhen Wu, Austin Marga, Lijie Zhang, Hao Zeng\*, Shaoming Huang\**

K. Zhang, Prof. S. Huang
Guangzhou Key Laboratory of Low-Dimensional Materials and Energy Storage Devices,
School of Materials and Energy,
Guangdong University of Technology,
Guangzhou 510006, China.
E-mail: smhuang@gdut.edu.cn

A. Marga, Prof. H. Zeng
Department of Physics,
University of Buffalo, Buffalo,
NY 14260, USA.
E-mail: haozeng@buffalo.edu

B. Pan, L. Zhang
Key Laboratory of Carbon Materials of Zhejiang Province,
Institute of New Materials and Industrial Technologies,
College of Chemistry and Materials Engineering,
Wenzhou University,
Wenzhou 325035, China.

C. Ding, Z. Wu
Key Laboratory of High-Performance Scientific Computation,
School of Science,
Xihua University,
Chengdu 610039, China.

**Abstract**

Understanding the growth mechanisms of two-dimensional (2D) van der Waals (vdW) heterostructures is of great importance in exploring their functionalities and device applications. A custom-built system integrating physical vapor deposition and optical microscopy/Raman spectroscopy was employed to study the dynamic growth processes of 2D vdW heterostructures *in situ*. This allows us to identify a new growth mode with a distinctly different growth rate and morphology from those of the conventional linear growth mode. We propose a model that explains the difference in morphologies and quantifies the growth rates of the two modes by taking the role of surface diffusion into account. We have systematically investigated a range of material combinations including $CdI_2/WS_2$, $CdI_2/MoS_2$, $CdI_2/WSe_2$, $PbI_2/WS_2$, $PbI_2/MoS_2$, $PbI_2/WSe_2$ and $Bi_2Se_3/WS_2$. These findings may be generalized to the synthesis of many other 2D heterostructures with controlled morphologies and physical properties, benefiting future device applications.

## 1. Introduction

Atomically thin 2D vdW heterostructures provide a new material platform for fundamental research [1-6] and applications in electronics, optoelectronics, spintronics and valleytronics, etc. [7-11] Most vdW heterostructures are obtained by transferring and stacking individual layers. However, using this method, it is challenging to precisely control the size, shape, layer thickness, and interface properties of the heterostructures, along with being ill-suited for the scale-up needed for device applications. [12-15] Although considerable work has been done in the past few years for the direct synthesis of vdW heterostructures using physical vapor deposition (PVD) and chemical vapor deposition (CVD), [16-19] the growth mechanisms are not well understood. [20-22] This is partly due to the fact that most studies on growth mechanisms relied on *ex situ* characterizations and often drew similarities to the growth of a single 2D vdW material (e.g. graphene or transition metal dichalcogenides, TMD) on non-vdW substrates. [23, 24] It has been found that in typical growth of vdW 2D materials, the 2D domains tend to evolve into regular shapes (e.g. triangular or hexagonal) under equilibrium conditions, following the Wulff construction (diffusion dominating over adhesion), [25-27] whereas they can yield dendritic or fractal shapes under conditions away from equilibrium. [28, 29]

*In situ* electron, atomic force and optical microscopies, and spectroscopies allow the capturing of the growth dynamics. [30-33] A range of 2D materials have been studied by these *in situ* techniques, including graphene, TMDs, organic crystals and MXenes. [34-38] Most of such studies focused on single materials grown on conventional metal or covalent substrates, and rarely have *in situ* imaging and spectroscopy been used on the investigation of epitaxial growth of 2D vdW heterostructures. For such heterostructures, a second 2D material nucleates and grows on the top surface of an existing vdW layer. Compared with conventional non-vdW substrates such as $SiO_2$, sapphire, and metals, the atomically smooth vdW surface lacks steps and dangling bonds. [39-42] The precursor molecules can diffuse nearly freely due to their weak interactions with the underlying vdW layer. This diffusion strongly modifies the growth dynamics of the vdW heterostructures. Therefore, clear differences exist in the growth mechanisms between vdW heterostructures and single 2D materials. The lack of detailed knowledge of growth dynamics may hinder the progress in the synthesis of novel 2D vdW heterostructures with precisely controlled morphology, stacking order, twisting angles, and thus their applications.

In this work, we report the development of a system that integrates physical vapor deposition with *in situ* optical microscopy/Raman spectroscopy, and the application of such technique to characterize the growth dynamics of 2D vdW heterostructures in real-time. We have grown a series of vdW heterostructures as prototype material systems, including $CdI_2/WS_2$, $CdI_2/MoS_2$, $CdI_2/WSe_2$, $PbI_2/WS_2$, $PbI_2/MoS_2$, and $PbI_2/WSe_2$ to demonstrate the universality of this system to reveal 2D heterostructure growth dynamics *in situ*. Our studies found that 2D materials preferentially nucleate on the vdW templates (rather than non-vdW substrates such as sapphire). We discovered two different growth modes which may be universal for 2D vdW heterostructures: one growth mode is characterized by its constant growth rate, resulting in regular shaped domains; the other has its growth rate inversely proportional to the domain size, leading to a suborbicular crystal shape. Using density functional theory (DFT) calculations, we replicated the two growth modes and explained their origins as the significant surface diffusion and edge adsorption effects. Based on this understanding, we designed experiments to synthesize other 2D vdW heterostructures (such as $Bi_2Se_3/WS_2$) using conventional CVD methods. By changing the synthesis parameters, we can control the growth in either mode. The different growth modes lead to significant differences in the growth rates, domain morphologies, and even interlayer twist angles of the 2D vdW heterostructures. Our results provide new insights into the growth dynamics of 2D vdW heterostructures, which is otherwise challenging to study using conventional methods.

## 2. Results and Discussion

### 2.1. *In situ*, real-time monitoring of the growth dynamics of 2D vdW heterostructures.

A system was built which integrates a temperature and environment-controlled stage for 2D material growth and a Raman microscope accessed through an optical window for optical microscopy and Raman spectroscopy/mapping for *in situ* monitoring of the growth dynamics (see Experimental Section). Monolayer $MX_2$ ($WS_2$, $MoS_2$, $WSe_2$) flakes were pre-grown on double-side polished sapphire substrates by conventional CVD. $YI_2$ ($CdI_2$, $PbI_2$) was thermally evaporated in the home-built system for growing $YI_2/MX_2$ heterostructures, as sketched in **Figure 1a** and Figure S1. The X-ray diffraction (XRD) patterns shown in Figure S2 indicate the grown $CdI_2$ and $PbI_2$ have high crystalline quality. The XRD pattern of $CdI_2$ matches well with standard PDF#12-0574, while that of $PbI_2$ matches with PDF#80-1000. Both are of hexagonal structure with a space group of $P\bar{3}m1$, and they show a strong (001) orientation expected for 2D $CdI_2$ and $PbI_2$. Six kinds of $YI_2/MX_2$ vdW heterostructures with different combinations of Y and M were fabricated and their growth processes were recorded in real-time.

Figure $1b_1$-$1g_1$ show the optical microscope images of the pre-grown single-layer TMD flakes on sapphire substrates. Figure $1b_2$-$1g_2$ show the images of as-grown $CdI_2/WS_2$, $CdI_2/MoS_2$, $CdI_2/WSe_2$, $PbI_2/WS_2$, $PbI_2/MoS_2$ and $PbI_2/WSe_2$ heterostructures, respectively. It can be seen that in all of these cases, isolated single microcrystals with triangular or hexagonal shapes are observed, thus reflecting the shapes of the underlying TMD flakes. Raman mapping is used to reveal the size and shape of individual $YI_2$ and TMD layers in the heterostructure. Raman mappings using the A mode ($A_1$ or $A_{1g}$) of $CdI_2$ or $PbI_2$ are shown in Figure $1b_3$-$1g_3$, and those with the $E_{2g}^1$ and E' modes of TMD are shown in Figure. $1b_4$-$1g_4$. It can be seen clearly that the two sets of images match well with each other, showing identical size and shape of $YI_2$ crystals with the underlaying monolayer TMDs. This suggests that the $YI_2$ crystals grew selectively on the vdW TMD monolayers and cover the entire flakes, while no material was deposited on the sapphire substrate. The typical Raman spectra of single-layer $MX_2$, bilayer $YI_2/MX_2$, and multilayer $YI_2/MX_2$ heterostructures are presented as black, red, and blue curves in Figure $1b_5$-$1g_5$, respectively. The Raman spectra of $YI_2/MX_2$ heterostructures not only demonstrate the characteristic peaks of $MX_2$, but also show those of $YI_2$ ($A_1$ of $CdI_2$; $E_{2g}$, $A_{1g}$, and 2LA(M) of $PbI_2$). Meanwhile, the 2LA(M) mode of $WS_2$, $A_{1g}$ mode of $MoS_2$, and $E_{2g}^1$ mode of $WSe_2$ soften by $\sim 1$–$1.5$ cm$^{-1}$ after the $YI_2$ was grown on the $MX_2$ layers, which can be attributed to the interlayer coupling between $YI_2$ and $MX_2$ layers, [43, 44] as shown in Figure S3.

## 2.2. Growth dynamics of the 2D vdW heterostructures.

*In-situ*, time-lapse optical microscope images reveal the growth dynamics and give deeper insights into the growth mechanisms of the 2D vdW heterostructures. **Figure 2a** shows the time-lapse images of the growth processes of $CdI_2/WS_2$ heterostructures at 285 °C (movies are shown in S1). For $CdI_2$ on $WS_2$ at 10 s, it can be seen that a nucleus was formed at the top right corner with no well-defined facets. With increasing time to 30 s, the nucleus grew and evolved into a hexagonal shaped crystal. With further increasing time, the domain grew isotropically in all directions, keeping the hexagonal shape unchanged. This process continued until the $CdI_2$ crystal reached one of the edges of $WS_2$. The edge shape of $CdI_2$ then conformed to that of $WS_2$, and the growth in that direction ceased, while the other edges of the $CdI_2$ crystal continued to grow. Interestingly, no growth of $CdI_2$ was observed beyond the boundaries of $WS_2$. In this particular example, a second $CdI_2$ nucleation event was observed at the lower-left corner at 30 s; the growth continued in a similar fashion. Atomic force microscopy (AFM) images clearly show that the growth edges remain flat and smooth during the growth process. The growth front can consist of either monolayer or step edges with different layer thicknesses, as shown in Figure 2b-2c. Figure 2f (tetragonal points) shows the size evolution of the $CdI_2$ domains as a function of time. By fitting this data, we can obtain a linear relationship between the domain size and the growth time. The growth rate $R$, as measured by the side length of the domains as a function of time, is found to be 0.41 µm/s for CdI$_2$. The Pearson correlation coefficients of the linear fitting are 0.995, indicating excellent linearity. Moreover, we found that multiple islands can nucleate on a single TMD flake and exhibit similar growth rates (Figure S4). We suggest that the number of nuclei on each TMD flake is related to the surface defect density, as the precursor molecules prefer to nucleate at energetically favorable defect sites. [45]

Surprisingly, by lowering the growth temperature, a different growth mode is observed (termed as sublinear growth mode). Figure 2d (a movie is shown in S2) shows the time-lapse images of the growth process of a CdI$_2$/WS$_2$ heterostructure at 260 ℃. It can be seen that the nucleated pattern has a suborbicular shape with no clearly defined facets. A magnified AFM image shows that the pattern exhibits irregular sawtooth edges instead of smooth ones (Figure 2e and Figure S5). Figure 2f gives a comparison of the size evolution process between sublinear and linear growth modes. Different from the linear growth mode that shows a constant growth rate, the size of the low temperature grown YI$_2$ increases with time in a sublinear fashion. This suggests that the growth rate decreases with time. The clear differences in the growth rates and morphologies between linear and sublinear modes suggest distinct growth mechanisms of the 2D vdW heterostructures that require further elucidation.

**2.3. Modeling of the growth mechanisms of 2D vdW heterostructures.**

The growth of YI$_2$ on TMD monolayers to form 2D vdW heterostructures is the result of a dynamic balance between precursor adsorption, desorption and surface diffusion. Calculating the energetics of these processes will help us to understand first and foremost, why YI$_2$ prefers two-dimensional over three-dimensional growth on TMDs, and second, why YI$_2$ prefers to grow on top of TMDs versus the sapphire surface. The adsorption energies for CdI$_2$ molecules over WS$_2$ (001), CdI$_2$ (001), and edges of CdI$_2$/WS$_2$ heterostructure were calculated by the Vienna *ab initio* simulation package (VASP). [46, 47] The adsorption energy for the CdI$_2$ molecules on the growth edges of the heterostructure (CdI$_2$/WS$_2$) is about 0.82 eV (**Figure 3a**).

However, the adsorption energies for $CdI_2$ molecules over $WS_2$ (001) and $CdI_2$ (001) are lower, at 0.3 eV and 0.55 eV, respectively (Figure 3b and 3c). It has been shown in recent work that there is strong fluctuation in the sizes of the nuclei at the early crystal nucleation stage, and a nucleus will only continue to grow when a critical size is reached. [48] For the growth of $YI_2$ on TMDs, once the nucleus reaches the critical size, the growth will proceed by the attachment of precursor molecules at the edges, as the rate of adsorption there far exceeds the rate of desorption. On the top surfaces of $YI_2$ and $WS_2$, due to the much lower adsorption energies, the rate of adsorption and desorption is balanced at the optimized growth temperatures, and thus a nucleus is unable to grow beyond the critical size and eventually completely evaporates. Therefore, the 2D growth mode of $YI_2$ is a consequence of the differential adsorption energies of the edges and top surfaces. The absence of growth on the sapphire substrate can be understood similarly. The adsorption energy of $CdI_2$ on sapphire is found to be 0.31 eV (Figure S6a). While the diffusion energy barrier is found to be 4.5 eV (Figure S6b). Thus, it is difficult for $CdI_2$ molecules to aggregate to form nuclei above the critical size for their further growth on sapphire. As shown by the images and movies, $YI_2$ crystals grow selectively on TMDs instead of on sapphire substrates.

With the understanding of the selective 2D growth, we next focus on the two observed growth modes with different time dependences at different temperatures. As has been discussed above, the 2D growth is dictated by initial nucleation followed by edge growth through the competition between adsorption and desorption of precursor molecules at growth edges. Precursor molecules can arrive at the growth edges mainly in two ways: first, precursor molecules can directly attach to the growth edges from the vapor phase; second, molecules can diffuse along the vdW surfaces and then attach to the growth edges (Figure 3e). At a relatively high temperature (higher than 285 °C), the weak adsorption energies of $CdI_2$ molecules on $CdI_2$ and $WS_2$ top surfaces (0.3 eV and 0.55 eV, respectively) are readily overcome by thermal energy, resulting in facile desorption of CdI₂ molecules from the surface of 2D layers. These precursor molecules thus do not contribute to the growth rate of $CdI_2$ 2D crystals. Since the adsorption energy of 0.82 eV at the growth edges is much larger than those at top surfaces, the $CdI_2$ molecules attach to the growth edges mainly by direct adsorption from the vapor phase. At sufficiently high growth temperatures, the edge shape is an outcome of the competition between adsorption and desorption. The morphology of the early stage nucleus likely possesses irregular sawtooth edges. [48, 49] These concave edges are thermodynamically favorable sites for adsorption due to the higher adsorption energy (see calculated values for Figure S7). The precursor molecules that go deep into the concave edge sites are also more likely to be trapped and adsorbed due to the increased probability of multiple collisions. On the other hand, the $CdI_2$ molecules adsorbed at the convex edge sites have a higher probability of desorption since the direction of diffusion is random, which decreases the chance of re-adsorption. [29] Hence, the concave sites are preferentially filled, resulting in flat edges (Figure 3f, red dash). The overall triangular or hexagonal shape of the 2D $CdI_2$ crystal is dictated by the underlying hexagonal crystal symmetry. Our calculation further shows that the edge shape is not determined by edge diffusion, as the energy barrier of $CdI_2$ molecule diffusion is prohibitively high (~1.25 eV as shown in Figure S8). The growth rate ($R$) of a 2D domain is the result of the competition between edge adsorption and desorption, which can be presented as: [50]

$$R = \frac{dL}{dt} = (c_1 f e^{-\frac{E_{ba}}{kT}} - c_2 f e^{-\frac{E_{bd}}{kT}}) s_0 \qquad (1)$$

where $c_1$ is the linear density of the precursor molecules in the vapor phase; $c_2$ is the linear density of the molecules adsorbed at the growth edge; $f = kT/h$, is the attachment/detachment frequency; $E_{ba}$ and $E_{bd}$ are energy barriers of source molecules/atoms attaching to the growth edge and detaching from the growth edge (not calculated here); $k$ is the Boltzmann constant; $h$ is the Planck constant; $T$ is the growth temperature; $s_0$ is the unit cell area of the 2D material; $L$ is the diameter of the 2D domain; and $t$ is the growth time. As mentioned earlier, at relatively high temperatures (higher than 285 °C), the $CdI_2$ molecules attach to the growth edges mainly from the vapor phase by direct adsorption. At equilibrium for a given temperature, both $c_1$ and $c_2$ are constants. Therefore, a constant $R$ of $CdI_2$ is expected because all the parameters in *Eq.* (1) are constants, resulting in a size dependence of the $CdI_2$ linear in time. Our setup allows for precise measurement of the growth rate of the 2D crystals. For $CdI_2$ crystals, $R$ is found to be in the range of 0.41-0.48 µm/s.

With decreasing growth temperatures (below 260 °C), the main difference is that the desorption rate of $CdI_2$ molecules landing on the top surface decreases. Meanwhile, the low diffusion energy barrier (0.15 eV on $WS_2$ and 0.22 eV on $CdI_2$) suggests fast diffusion of $CdI_2$ molecules on $WS_2$ and $CdI_2$ layers (Figure 3d). Therefore, the $CdI_2$ molecules within the mean-free-path of the edges can arrive and get adsorbed at the edges. (Figure 3g). This results in a growth rate that is time-dependent, which will be derived below. The second difference is that the desorption rate of $CdI_2$ molecules at the growth edges also decreases dramatically with decreasing temperature. Therefore, the edge shape is kinetically controlled. This leads to a suborbicular shape of the as-grown $CdI_2$ 2D crystals since the surface diffusion is isotropic. The edge shapes are irregular due to the fast growth rate, with the edge voids unable to be filled before new growth.

Compared with the linear mode, the sublinear growth mode exhibits a higher growth rate due to the higher precursor density near the growth edges and decreased desorption rate. The concentration of precursor $CdI_2$ molecules ($c$) near the growth edge is now composed of three parts: the $CdI_2$ molecules in the vapor phase ($c_1$), and those adsorbed on the top surfaces of $WS_2$ and $CdI_2$ within the mean free paths $\lambda_1$ and $\lambda_2$, respectively ($c_S$). Assume the growing $CdI_2$ to be of a circular shape, and the $CdI_2$ molecules adsorbed on the surfaces of $WS_2$ and $CdI_2$ within the mean free paths to the edges (of the shape of two circular rings with widths of $\lambda_1$ and $\lambda_2$, respectively, as shown in Figure S9) can diffuse and get adsorbed at the edges, $c$ can be presented as:

$$c = c_1 + c_S = c_1 + \frac{c_{S1}S_1 + c_{S2}S_2}{\pi L} = c_1 + c_{S1}\lambda_1 + c_{S2}\lambda_2 + \frac{c_{S1}\lambda_1^2 - c_{S2}\lambda_2^2}{L} \qquad (2)$$

where $c_{S1}$ and $c_{S2}$ are the areal concentrations of the precursor molecules adsorbed on $WS_2$ and $CdI_2$ layers, respectively; $S_1$ and $S_2$ are the effective adsorption areas for $CdI_2$ molecules over $WS_2$ and $CdI_2$ layers within the mean free paths $\lambda_1$ and $\lambda_2$, respectively; and $L$ is the diameter of the $CdI_2$ layer.

The explicit expression of $\frac{dL}{dt}$ is obtained by substituting the linear concentration of precursor $c$ in Eq. (1) by Eq. (2):

$$\frac{dL}{dt} = \frac{A}{L} + B \qquad (3)$$

where $A = (c_{S1}\lambda_1^2 - c_{S2}\lambda_2^2)s_0 f e^{-\frac{E_{ba}}{kT}}$ and $B = \left[ (c_1 + c_{S1}\lambda_1 + c_{S2}\lambda_2)e^{-\frac{E_{ba}}{kT}} - c_2 e^{-\frac{E_{bd}}{kT}} \right] s_0 f$ are both constants independent of time. Eq. (3) shows that the size of the $CdI_2$ domains as a function of growth time exhibits a sublinear behavior. This sublinear behavior originates from the significant contribution of precursor molecules that arrive at growth edges via surface diffusion. The fitting curves by Eq. (3) agree well with the experimental data, as shown in Figure 2f (pink line).

## 2.4. Universality of the growth models.

While we used $CdI_2/WS_2$ to illustrate the two different growth models at different growth temperatures, these models should be generally applicable to other vdW heterostructures. Based on the understanding of these two models, we designed experiments to synthesize other 2D vdW heterostructures with different morphologies, including $CdI_2$ and $PbI_2$ on different TMDs. It is found that $PbI_2$ follows the behavior of $CdI_2$ closely, with linear and sublinear growth modes at high and low temperatures, respectively. The critical temperature separating the two modes is found to be 330 °C. As shown in **Figure 4a**, 4b, the suborbicular and triangular PbI$_2$/WS$_2$ can be controllably grown at 300 and 330 °C, respectively. As expected, the sublinear growth mode of PbI$_2$ exhibits a higher growth rate, as shown in Figure 4c. Furthermore, the sublinear growth mode always arises regardless of the underlying vdW layers, as long as the growth temperature is sufficiently low. Figure 4d and 4e show the time-lapse images of the sublinear growth processes of CdI$_2$/MoS$_2$ and CdI$_2$/WSe$_2$ heterostructures at 260 °C (movies shown in S3 and S4). The optical microscope images of completed CdI$_2$/MoS$_2$ and CdI$_2$/WSe$_2$ heterostructures (growth time of ~320 s and 95 s, respectively) are shown in Figure 4f and 4g. As shown in Figure 4h, the crystal size as a function of time for CdI$_2$/WS$_2$, CdI$_2$/MoS$_2$, and CdI$_2$/WSe$_2$ heterostructures can all be fitted by Eq. (3). Moreover, CdI$_2$ has the highest growth rate on WS$_2$ compared to that on MoS$_2$ and WSe$_2$, due to the smallest diffusion barrier of WS$_2$ (see Figure 3d and Figure S10).

We further synthesized 2D Bi$_2$Se$_3$, a topological insulator, with different morphologies, on WS$_2$ using conventional CVD methods. **Figure 5a-5d** show the AFM and Raman mapping of Bi$_2$Se$_3$/WS$_2$ vdW heterostructures grown at 600 °C and 505 °C, respectively, with identical growth time. It can be seen that the shapes of Bi$_2$Se$_3$ grown at 600 °C are triangular and hexagonal, with smooth edges; whereas those grown at 505 °C are suborbicular and possess irregular sawtooth edges (Figure 5a, 5c and Figure S11). Although both are grown for the same amount of time, the domain size of Bi$_2$Se$_3$ grown at 505 °C is clearly larger than that generated at 600 °C (Figure 5a, 5c and Figure S11). This confirms the model predictions convincingly, even though the growth rates cannot be easily obtained in conventional CVD growth. To gain deeper insights into the differences between the 2D vdW heterostructures grown by the two modes, we used transmission electron microscopy (TEM) and selected area electron diffraction (SAED) to investigate possible epitaxial relationship between Bi$_2$Se$_3$ and WS$_2$. Figure 5e-5n depict the low-magnification TEM images and SAED patterns of Bi$_2$Se$_3$/WS$_2$ heterostructure samples grown by both modes. The SAED shows two different sets of hexagonal diffraction patterns (highlighted by green and red dashed lines), corresponding to the $WS_2$ lattice (0.27 nm) and the $Bi_2Se_3$ lattice (0.21 nm), respectively (Figure S12). The two sets of SAED patterns are clearly aligned for the sample grown at 600 °C, as shown in Figure 5e and 5f. A total of 15 flakes were measured and 100% of them are aligned. On the other hand, for $Bi_2Se_3$ grown at 505 °C, while majority of the flakes (17 out of 20 tested) show SAED patterns aligned to that of $WS_2$, different twisting angles including 1.1°, 3.7°, and 30° are also observed (Figure 5g-5n and S13). This can be understood as following: the higher growth temperature provides sufficient thermal energy for the $Bi_2Se_3$ islands initially nucleated on the $WS_2$ layer to relax into the most thermodynamically stable configuration. However, with decreasing growth temperature, the initially nucleated $Bi_2Se_3$ island with a twisted angle cannot overcome the high diffusion barrier to reorient itself. The interlayer twisted angle is an important parameter for 2D vdW heterostructures because it may create flat bands with localized states and enhanced electronic correlations that can lead to exotic many body physics such as superconductivity and ferromagnetism. [51-54] To date, most of the heterostructures with twisting angles were obtained by transferring and layer stacking, which could lead to potential issues of interface contamination. We suggest that our sublinear growth mode may provide a convenient method to grow 2D vdW heterostructures with variable interlayer twisting angles, allowing the investigation of new many body physics.

## 3. Conclusions

In conclusion, the dynamic growth processes of several 2D vdW heterostructures were systematically investigated by *in situ* imaging and Raman spectroscopy using a custom-built system. While a commonly known linear growth mode with a constant growth rate is confirmed by *in situ* studies, a new growth mode, namely the sublinear mode, is also identified. The two modes demonstrate substantially different growth rates and morphologies. This difference is mainly attributed to the contribution of surface diffusion of adsorbed precursor molecules to the edge growth, which can be controlled by substrate temperature. While a limited number of vdW layer combinations were studied, it is expected that these models are applicable to the growth processes of a broad range of vdW heterostructures.

## 4. Experimental Section

*Preparation of monolayer $WS_2$, $MoS_2$, and $WSe_2$:* The sapphire substrates (0.1 mm, double-sided polished) were placed in the center of a quartz tube furnace, and the $WO_3$ (or $MoO_3$) powder was dispersed on the sapphire substrates. $Ar/H_2S$ with a flow rate of 30/10 sccm was used as the carrier gas and S source. The furnace was heated to 1000 °C (or 800 °C) and was held at that temperature for 2 min for the growth of monolayer $WS_2$ (or $MoS_2$). The $WO_3$ powder in an alumina boat was placed in the center of the quartz tube furnace, and sapphire substrates (0.1 mm, double-sided polished) were placed face down on top of the alumina boat. Another boat containing Se powder was placed further upstream at 300 °C. $Ar/H_2$ with a flow rate of 35/5 sccm was used as the carrier gas. The furnace was heated to 900 °C and was held at that temperature for 2 min for growing $WSe_2$.

*Growth of $CdI_2/WS_2$, $CdI_2/MoS_2$, $CdI_2/WSe_2$, $PbI_2/WS_2$, $PbI_2/MoS_2$, $PbI_2/WSe_2$ 2D vdW heterostructures:* The custom-made system consists of a temperature and environment control stage (Linkam THMS600) and *in situ* imaging system (invia Renishaw micro-Raman). The growth apparatus consists of a heating stage and a water-cooler. An optical window made of quartz was used for *in situ* optical imaging and spectrum collection. The system was purged with high-purity Ar gas during the growth process. 1 mg precursor powder ($CdI_2$ or $PbI_2$) was dispersed on the heating stage. A sapphire substrate (with pre-grown TMD) was placed face down on top of the heating stage. The spacing between the substrate and the heating stage was set at 500 μm. The growth temperature was as follows: $CdI_2/TMD$ heterostructures (260 °C for sublinear growth mode; 285 °C for linear growth mode), $PbI_2$/TMD heterostructures (300 °C for sublinear growth mode; 330 °C for linear growth mode).

*Growth of $Bi_2Se_3$/$WS_2$ 2D vdW heterostructure:* The $Bi_2Se_3$/$WS_2$ heterostructure was grown by a CVD tube furnace. $Bi_2Se_3$ powder was placed in a quartz boat in the tube furnace. A sapphire substrate (with pre-grown $WS_2$) was placed downstream near the $Bi_2Se_3$ powder. The quartz tube was kept in an argon atmosphere for 40 min to remove oxygen. The furnace was heated to the target temperature (505 °C-600 °C) for the growth of $Bi_2Se_3$/$WS_2$ heterostructures (505 °C for sublinear growth mode; 600 °C for linear growth mode).

*Characterizations: In situ* Raman measurements were performed using an invia Renishaw micro-Raman system with a 532 nm excitation laser. Raman spectra were calibrated by the Raman shift of single-crystal silicon at 520.4 cm$^{-1}$. The thicknesses of 2D heterostructures were measured by AFM (Dimension Icon, Bruker). TEM measurements were conducted in a JEOL 2100F. X-ray diffraction (XRD) patterns were recorded on a Bruker D8-Advance system.

*DFT Calculations:* All first-principles calculations were performed using the Vienna *ab initio* simulation package (VASP) based on density functional theory (DFT). The frozen-core projector augmented wave (PAW) method was adopted to describe the interaction between the core and valence electrons and the Perdew-Burke-Ernzerhof (PBE) exchange correlation functional was used to depict the interactions within the generalized gradient approximation (GGA) framework. The vdW interactions for the $CdI_2$/$WS_2$ and $Bi_2Se_3$/$WS_2$ heterostructures were considered by adopting Grimmes DFT-D3 method. In all calculations, 7×7×1 K-points were selected for the Monkhorst-Pack mesh to sample the 2D Brillouin Zone. The cutoff energy for the plane wave adopted in the present work was 500 eV. In the structural relaxation and other calculations, the convergence criteria were chosen to be < 10$^{-5}$ eV for the total energy and 0.01 eV/Å for the force on each atom. After testing, a vacuum zone of 15 Å was employed to prevent the interactions between the adjacent images.

# WILEY-VCH

**Supporting Information**
Supporting Information is available from the Wiley Online Library or from the author.

**Acknowledgements**
We thank Shao-Yi Wu of University of Electronic Science and Technology of China for his help with the VASP calculations. Funding: This research was supported by the National Natural Science Foundation of China (52002080, 51920105004, 52072272), China Postdoctoral Science Foundation (2020M682610) and the Major Research Plan of Wenzhou City (ZG2017027). Author contributions: K.Z., L.Z. and S.H. designed the research. K.Z., and B.P. synthesized the materials and conducted optical microscope, AFM, XRD, Raman, TEM and data analysis. C.D. carried out the first-principles calculation. K.Z. and H.Z. proposed the growth models and analyzed the data. K.Z., W.Z., A. M., H.Z. and S.H. wrote the paper with input from other coauthors. The data that support the findings of this study are available from the corresponding author upon reasonable request.

Received: ((will be filled in by the editorial staff))
Revised: ((will be filled in by the editorial staff))
Published online: ((will be filled in by the editorial staff))

**Conflict of Interest**
The authors declare that they have no competing interests.

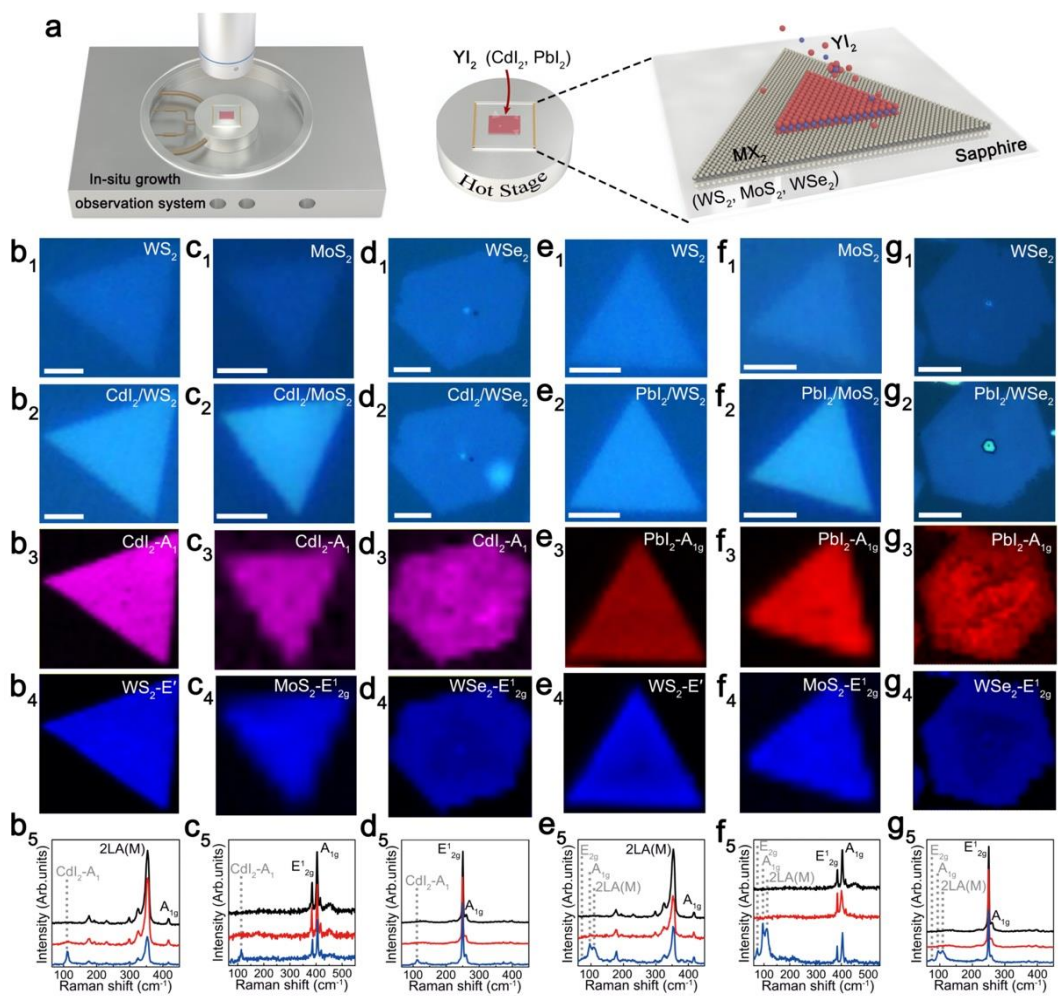

**Figure 1.** Growth and *in situ* characterizations of 2D vdW heterostructures. a) A schematic of the custom-made system for combined growth and *in situ* imaging of 2D $YI_2/MX_2$ vdW heterostructures. $b_1$, $b_2$) Optical microscopy images of pre-grown monolayer $WS_2$ and as-grown $CdI_2/WS_2$ heterostructure. $b_3$, $b_4$) *In situ* Raman peak intensity mappings of the $CdI_2$-$A_1$ mode and $WS_2$-$E'$ mode. $b_5$) Representative Raman spectra of $CdI_2/WS_2$ heterostructures. The black, red and blue lines represent the monolayer $WS_2$, bilayer $CdI_2/WS_2$ and multilayer $CdI_2/WS_2$ heterostructures. $c_1$-$g_1$) Optical microscopy images of pre-grown monolayer $MoS_2$, $WS_2$ and $WSe_2$. $c_2$-$g_2$) Optical microscopy images of as-grown $CdI_2/MoS_2$, $CdI_2/WSe_2$, $PbI_2/WS_2$, $PbI_2/MoS_2$ and $PbI_2/WSe_2$ heterostructures, respectively. $c_3$-$g_3$) *In situ* Raman peak intensity mappings of $CdI_2$-$A_1$ mode and $PbI_2$-$A_{1g}$ mode, corresponding to $c_2$-$g_2$. $c_4$-$g_4$) *In situ* Raman peak intensity mappings of $MX_2$-E mode, corresponding to $c_2$-$g_2$. $c_5$-$g_5$) Representative Raman spectra of $CdI_2/MoS_2$, $CdI_2/WSe_2$, $PbI_2/WS_2$, $PbI_2/MoS_2$, and $PbI_2/WSe_2$ heterostructures. The black, red and blue lines represent monolayer $MX_2$, bilayer $YI_2/MX_2$ and multilayer $YI_2/MX_2$ vdW heterostructures. Scale bars are 5 μm.

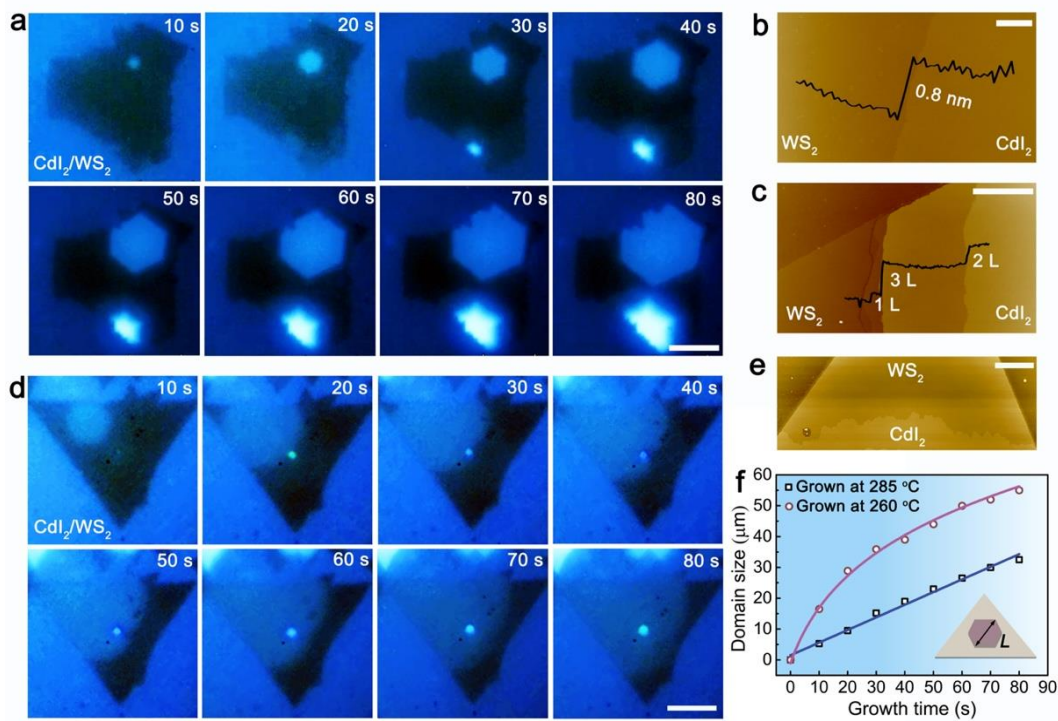

**Figure 2.** Growth dynamics of 2D $CdI_2/WS_2$ vdW heterostructures. a) *In situ* real-time images of the dynamic growth process of $CdI_2/WS_2$ heterostructure at the growth temperature of 285 °C. b, c) The AFM images of the boundaries of $CdI_2/WS_2$ heterostructure samples grown at 285 °C. d) *In situ* real-time images of the dynamic growth process of $CdI_2/WS_2$ heterostructure at the growth temperature of 260 °C. e) The AFM image showing the morphology of the boundaries of $CdI_2/WS_2$ heterostructure grown at 260 °C. f) Domain size as a function of time for $CdI_2/WS_2$ heterostructures grown at 285 and 260 °C, respectively. The dots are measured data and the lines are fitting curves using Eq. (1) and (3), respectively. Scale bars: 20 μm (a, d); 2 μm (b, c, e).

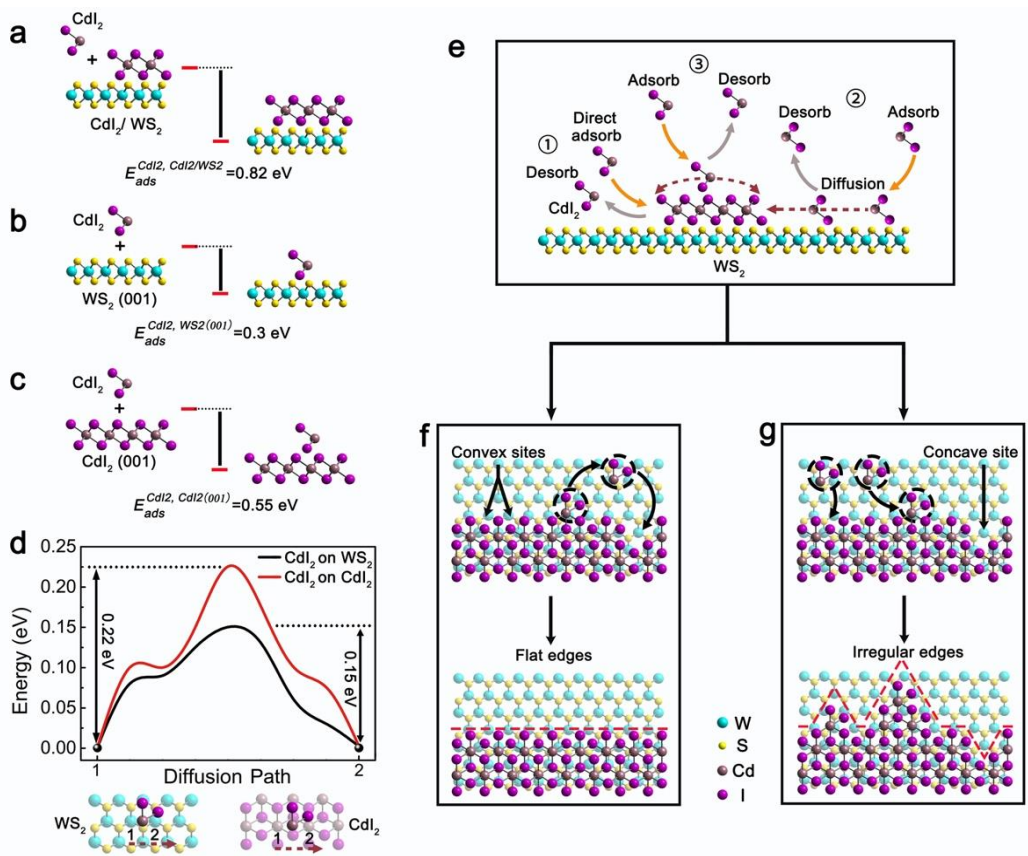

**Figure 3.** Theoretical analysis of the growth mechanisms of 2D vdW heterostructures. a-c) The calculated adsorption energy for $CdI_2$ molecules on $CdI_2/WS_2$ growth edges, over $WS_2$ (001) surface and over $CdI_2$ (001) surface, respectively. d) The calculated surface diffusion barriers for $CdI_2$ molecules on $WS_2$ (001) and $CdI_2$ (001) surfaces. Insets show the atomic models of simulated surface diffusion paths for $CdI_2$ molecules over $WS_2$ and $CdI_2$ layers respectively. e) A schematic showing the three routes for the $CdI_2$ molecules to arrive at the heterostructure growth edges: ① by direct attachment to the growth edges from vapor phase; by diffusion on ② $WS_2$ and ③ $CdI_2$ surface, respectively. f, g) Schematic growth processes of $CdI_2/WS_2$ vdW heterostructures for the linear and sublinear modes, respectively. The flat and sawtooth edges are highlighted by red dashed lines.

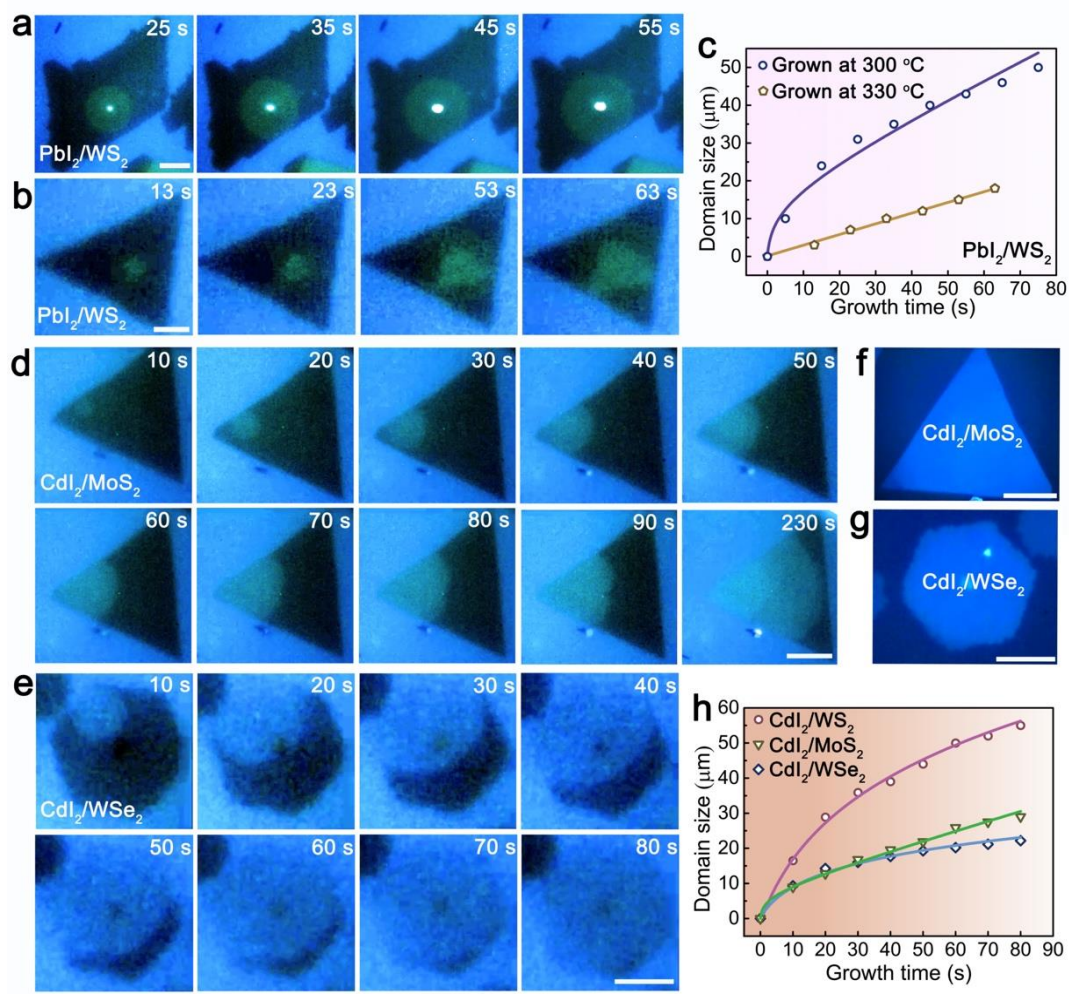

**Figure 4.** Dynamic growth process by *in situ* imaging of other 2D vdW heterostructures. a, b) *In situ* real-time images of growth dynamics of PbI₂/WS₂ heterostructure at temperatures of 300 °C and 330 °C, respectively. c) Domain size as a function of time for PbI₂/WS₂ heterostructures grown at 300 °C and 330 °C, respectively. d, e) *In situ* real-time imaging of growth dynamics of CdI₂/MoS₂ and CdI₂/WSe₂ heterostructures at the growth temperature of 260 °C. f, g) Optical microscope images of completed CdI₂/MoS₂ and CdI₂/WSe₂ heterostructures with growth times of 320 s and 95 s in panel d and e, respectively. h) Domain size of CdI₂/WS₂, CdI₂/MoS₂ and CdI₂/WSe₂ heterostructures as a function of time grown at 260 °C. The solid lines in c and h represent the fitting curves. Scale bars: 20 μm (a, d, f); 10 μm (b, e, g).

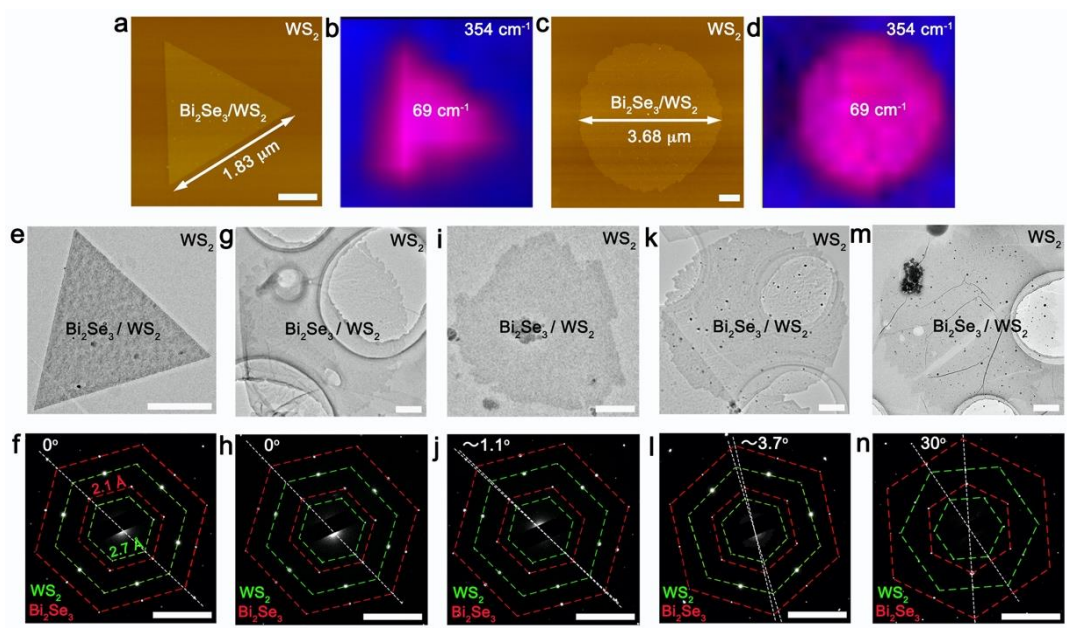

**Figure 5.** Characterizations of 2D Bi$_2$Se$_3$/WS$_2$ vdW heterostructures. a, b) The AFM images and Raman intensity mapping of Bi$_2$Se$_3$/WS$_2$ triangular shaped heterostructure grown at 600 °C. c, d) The AFM images and Raman intensity mapping of suborbicular shaped Bi$_2$Se$_3$/WS$_2$ heterostructure grown at 505 °C. e, f) The Low-magnification TEM image and SAED pattern of triangular shaped Bi$_2$Se$_3$/WS$_2$ heterostructure grown at 600 °C. g-n) The Low-magnification TEM images and SAED patterns of suborbicular shaped Bi$_2$Se$_3$/WS$_2$ heterostructures grown at 505 °C. Scale bars: 500 nm (a, c, e, g, i, k, m); 5 nm$^{-1}$ (f, h, j, l, n).

# Supporting Information

## Visualizing van der Waals Epitaxial Growth of Two-Dimensional Heterostructures

*Kenan Zhang, Changchun Ding, Baojun Pan, Zhen Wu, Austin Marga, Lijie Zhang, Hao Zeng\*, Shaoming Huang\**

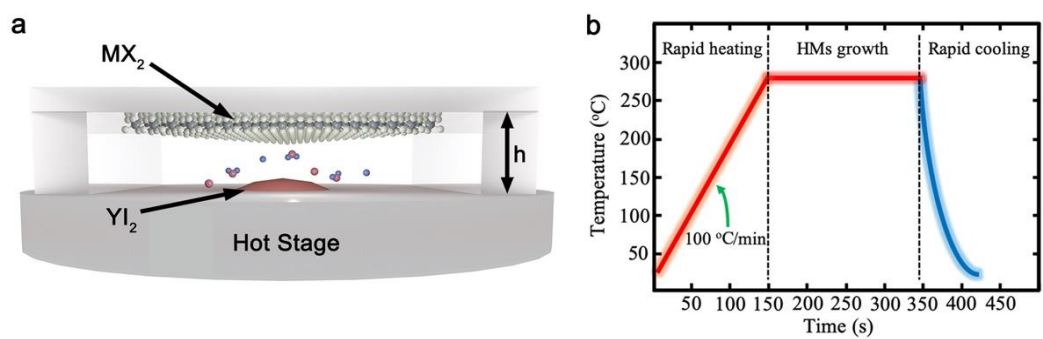

**Figure S1.** a) A schematic view the of the growth system. b) The time vs temperature profile for the growth of 2D $YI_2/MX_2$ heterostructures.

The Rayleigh number ($R_a$) [1, 2] is used to characterize the vapor and heat transport regimes and it is defined as:

$$R_a = \frac{g\beta\Delta T h^3}{\nu\kappa} \qquad (S1)$$

where $g$ is the gravitational constant, $\beta$ is the thermal expansion coefficient, $\Delta T$ is the temperature difference, $h$ is the distance between the source and deposition substrate, $\nu$ is the kinematic viscosity, and $\kappa$ is the thermal diffusivity. The Rayleigh number is estimated to be $0.7\sim1 \times 10^{-4}$ for the conditions used in this work, which indicates a buoyancy-driven vapor flow transport mode in the home-built growth system.

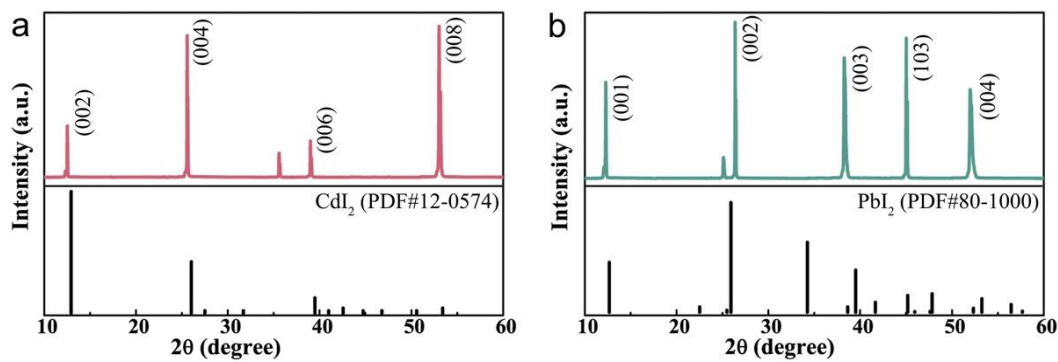

**Figure S2.** The XRD patterns of as-grown a) $CdI_2$ and b) $PbI_2$, respectively.

Bulk $CdI_2$ and $PbI_2$ were deposited on mica substrates using the home-built growth system. 10 mg of the precursor powder ($CdI_2$ or $PbI_2$) was put on the hot stage. The mica substrates were placed face down on top of the hot stage. Then the hot stage was heated to the desired growth temperature (285 °C for $CdI_2$ and 330 °C for $PbI_2$) and was held at that temperature for 30 minutes.

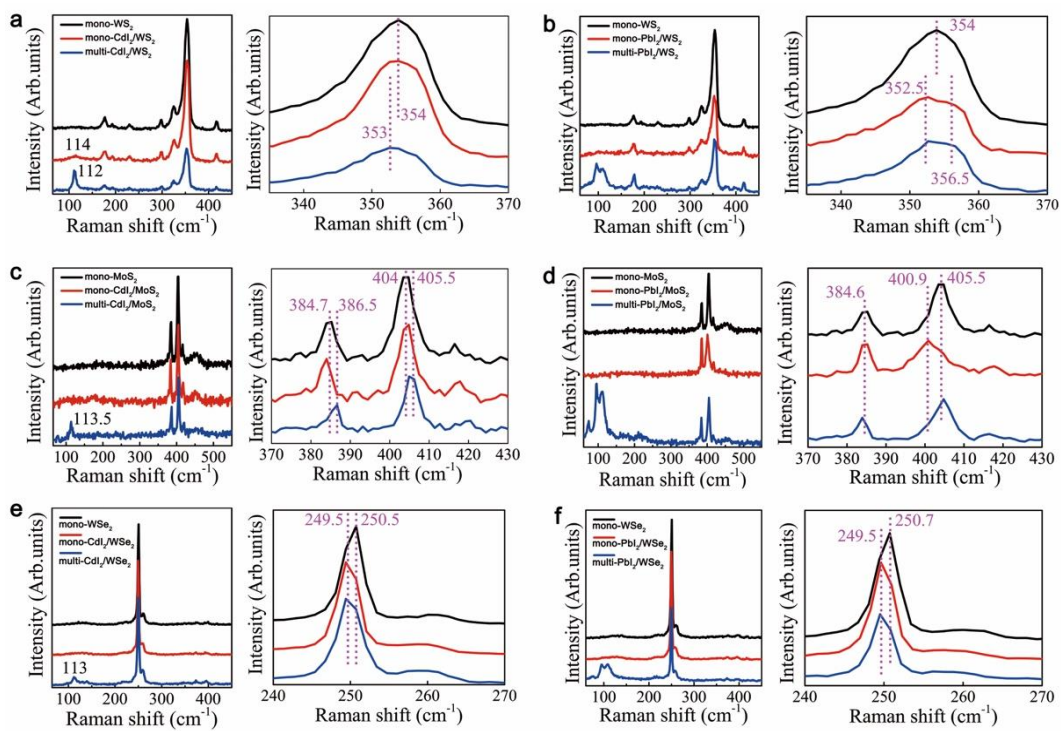

**Figure S3.** Raman spectra of 2D YI$_2$/MX$_2$ vdW heterostructures. Raman spectra of a) CdI$_2$/WS$_2$, b) PbI$_2$/WS$_2$, c) CdI$_2$/MoS$_2$, d) PbI$_2$/MoS$_2$, e) CdI$_2$/WSe$_2$ and f) PbI$_2$/WSe$_2$ heterostructures, respectively.

The MX$_2$ Raman peaks shift to lower wavenumbers by about $1{-}1.5$ cm$^{-1}$ after the YI$_2$ were grown on the MX$_2$ layers, which indicates the interlayer coupling between YI$_2$ and MX$_2$ layers.[3,4]

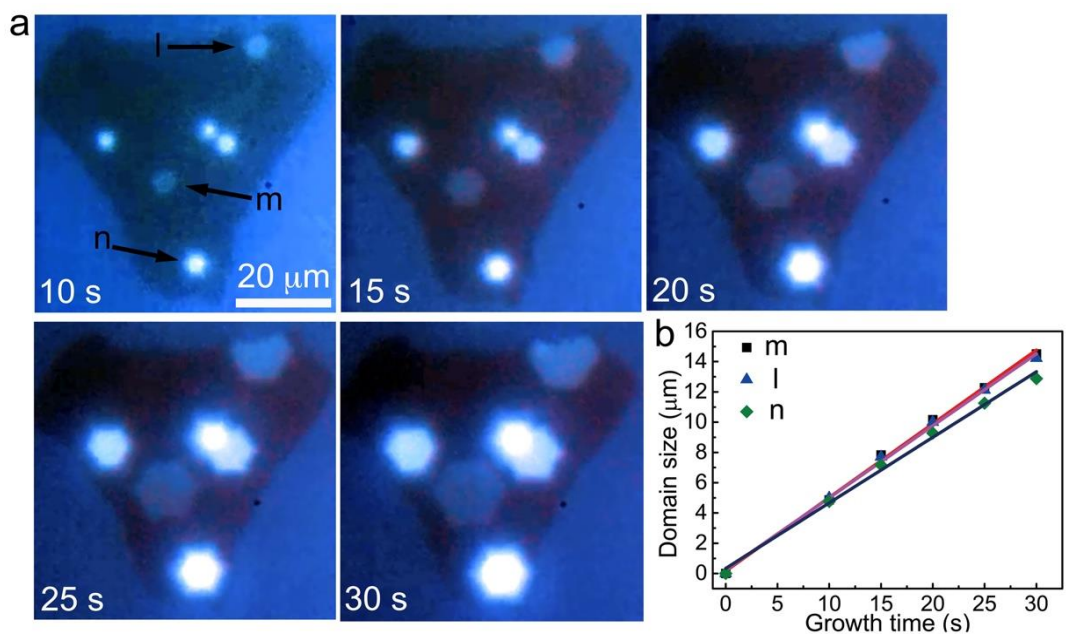

**Figure S4.** Simultaneous nucleation and growth of multiple flakes of the CdI$_2$/WS$_2$ heterostructure. a) *In situ* real-time images of the dynamic growth process of multiple flakes at the growth temperature of 285 °C. b) Domain size as a function of time for three CdI$_2$ islands (labeled as l, m and n in panel a).

The solid lines in panel b are the fitting curves by using Eq. (1).

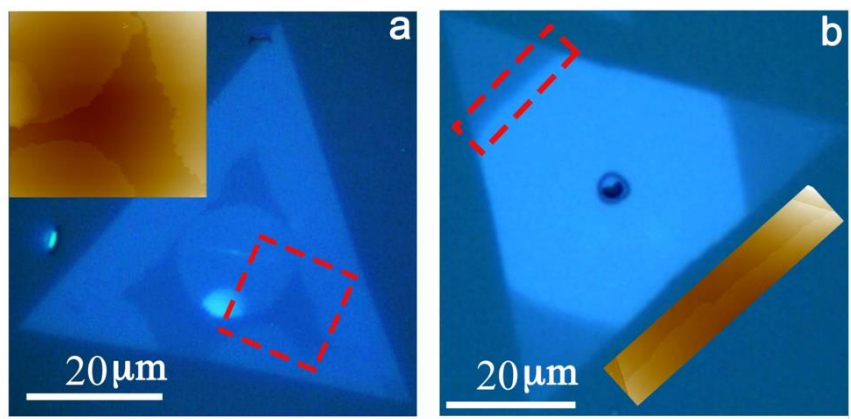

**Figure S5.** *Ex situ* AFM images of CdI$_2$/WS$_2$ growth edges. a) The growth edges of CdI$_2$/WS$_2$ heterostructures were irregularly sawtooth shaped when grown at 260 °C. b) While the growth edges are smooth at temperatures of 285 °C. Insets are AFM images corresponding to the labeled regions in optical microscope images.

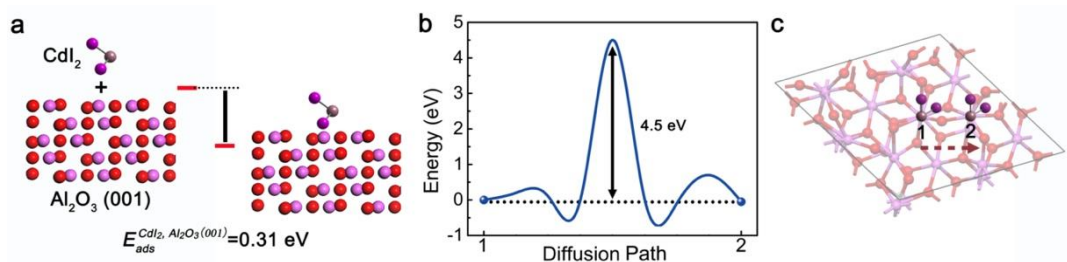

**Figure S6.** The simulated surface adsorption energy and diffusion barrier for CdI$_2$ molecules over the sapphire (001) substrate. a) Sideview of atomic models of CdI$_2$ molecules over sapphire substrates. The calculated surface adsorption energy for CdI$_2$ molecules over a sapphire substrate is about 0.31 eV. b) The calculated surface diffusion barrier for CdI$_2$ molecules on the sapphire (001) substrates is 4.5 eV. c) Atomic model of the simulated surface diffusion path for CdI$_2$ molecules on the sapphire substrate.

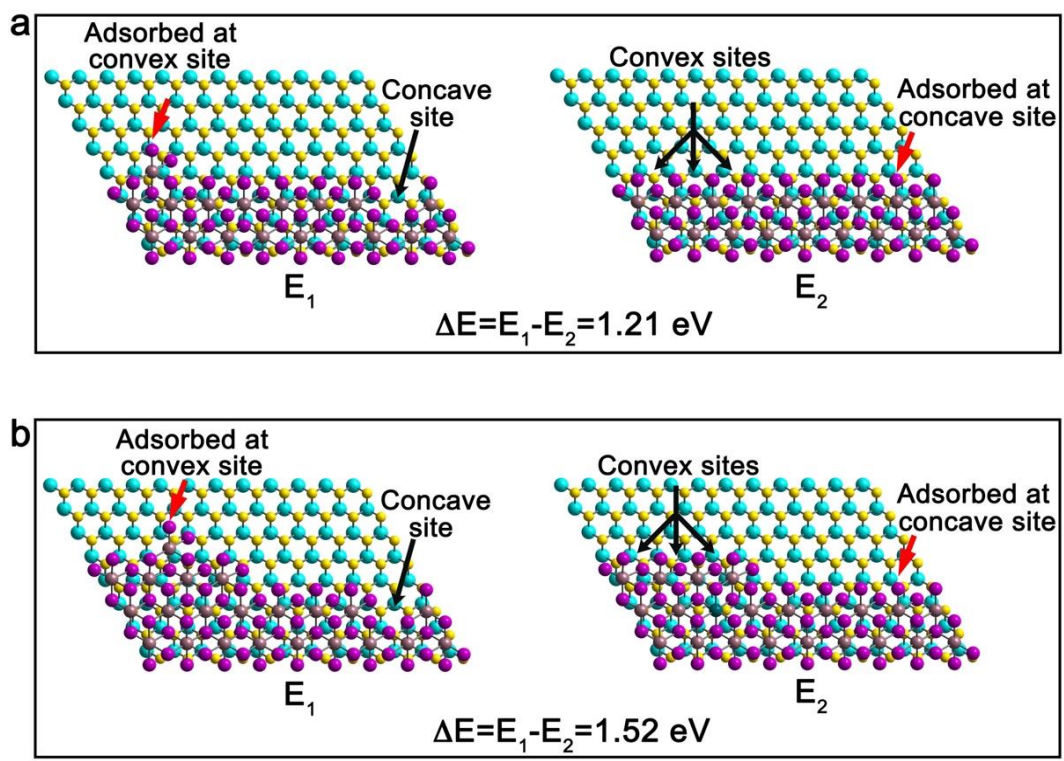

**Figure S7.** The simulated edge energy of the CdI$_2$/WS$_2$ heterostructure. a, b) Calculated energy difference ($\Delta E$) of CdI$_2$ molecules adsorbed at different heterostructure edge sites.

The positive $\Delta E$ values indicate concave edge sites are energetically favorable sites for precursor molecules compared with the convex sites.

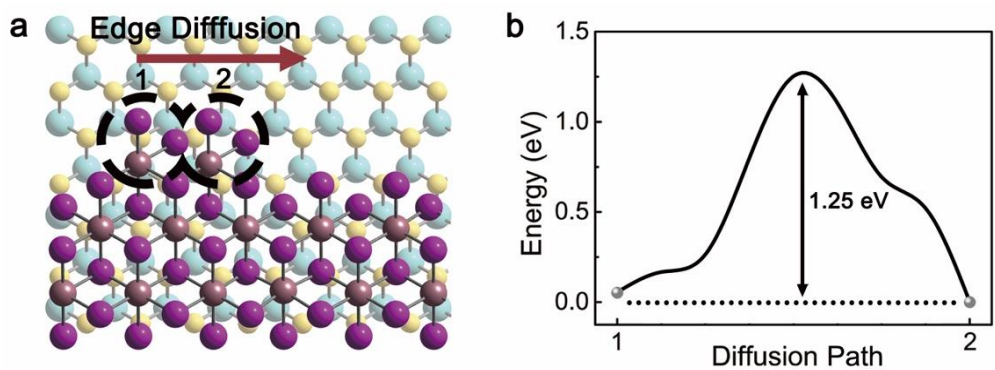

**Figure S8.** The simulated edge diffusion barrier for $CdI_2$ molecules on the edge of $CdI_2/WS_2$ heterostructure. a) Atomic model of simulated edge diffusion path for $CdI_2$ molecules on edges of $CdI_2/WS_2$ heterostructure. b) The calculated edge diffusion barrier for $CdI_2$ molecules on edges of $CdI_2/WS_2$ heterostructure is about 1.25 eV.

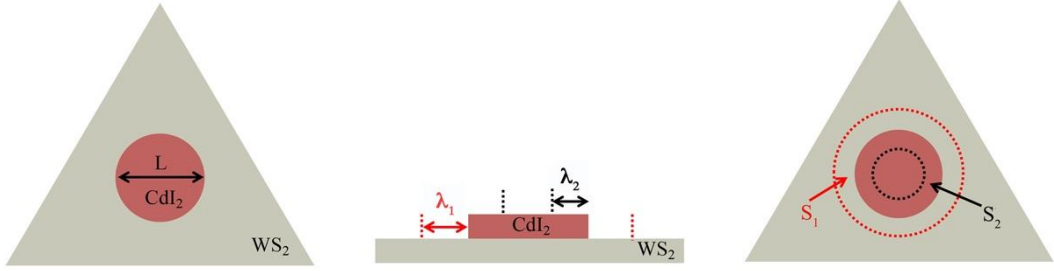

**Figure S9.** A schematic of the surface diffusion in the model of sublinear growth of the $CdI_2/WS_2$ heterostructure.

The edge shape of $CdI_2$ is kinetically controlled at low growth temperatures. This leads to a suborbicular shape of the as-grown $CdI_2$ 2D crystals, since the surface diffusion is isotropic. Here, we approximate the shape of the $CdI_2$ flake as a circle. $S_1$ and $S_2$ are the effective adsorption areas for $CdI_2$ molecules over $WS_2$ and $CdI_2$ layers within the mean free paths $\lambda_1$ and $\lambda_2$, respectively. $L$ is the diameter of $CdI_2$ layer.

$$S_1 = \left[\pi \left(\frac{L}{2} + \lambda_1\right)^2 - \pi \left(\frac{L}{2}\right)^2\right] = \pi L \lambda_1 + \pi {\lambda_1}^2 \text{ and}$$

$$S_2 = \left[\pi \left(\frac{L}{2}\right)^2 - \pi \left(\frac{L}{2} - \lambda_2\right)^2\right] = \pi L \lambda_2 - \pi {\lambda_2}^2$$

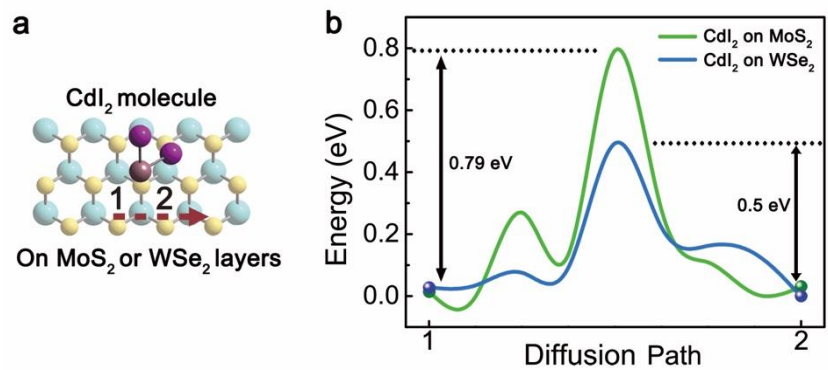

**Figure S10.** The simulated surface diffusion barriers for CdI$_2$ molecules on monolayer MoS$_2$ and WSe$_2$. a) Atomic model of the simulated surface diffusion paths for CdI$_2$ molecules on MoS$_2$ and WSe$_2$ layers. b) The calculated surface diffusion barriers of CdI$_2$ molecules on monolayer MoS$_2$ and WSe$_2$ layers are 0.79 eV and 0.5 eV respectively. So, CdI$_2$ has the smallest diffusion barrier on WS$_2$ (0.3 eV) compared with that on MoS$_2$ and WSe$_2$.

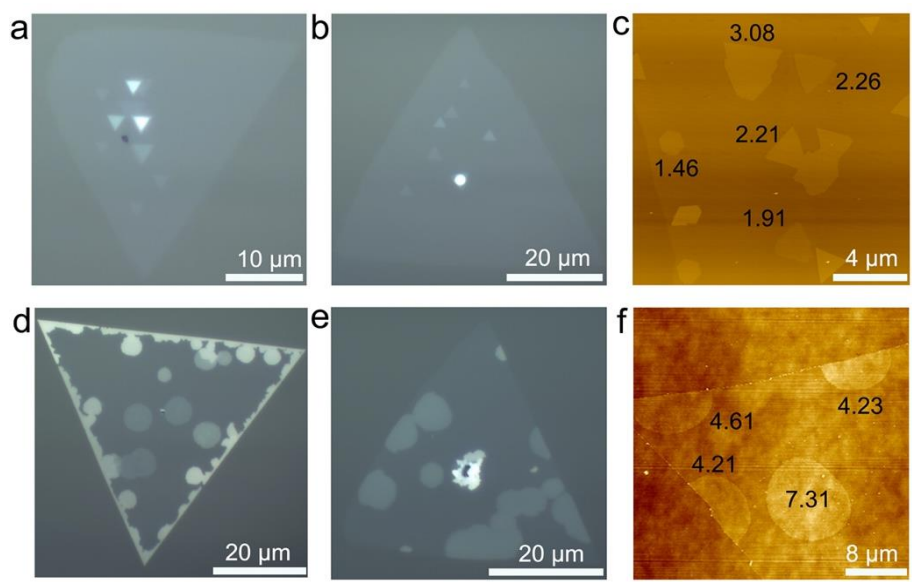

**Figure S11.** The Optical microscope and AFM images of 2D $Bi_2Se_3$/$WS_2$ vdW heterostructures grown at 600 °C a-c) and 505 °C d-f), respectively. The numbers shown in panel c and f are the measured domains sizes (unit: μm).

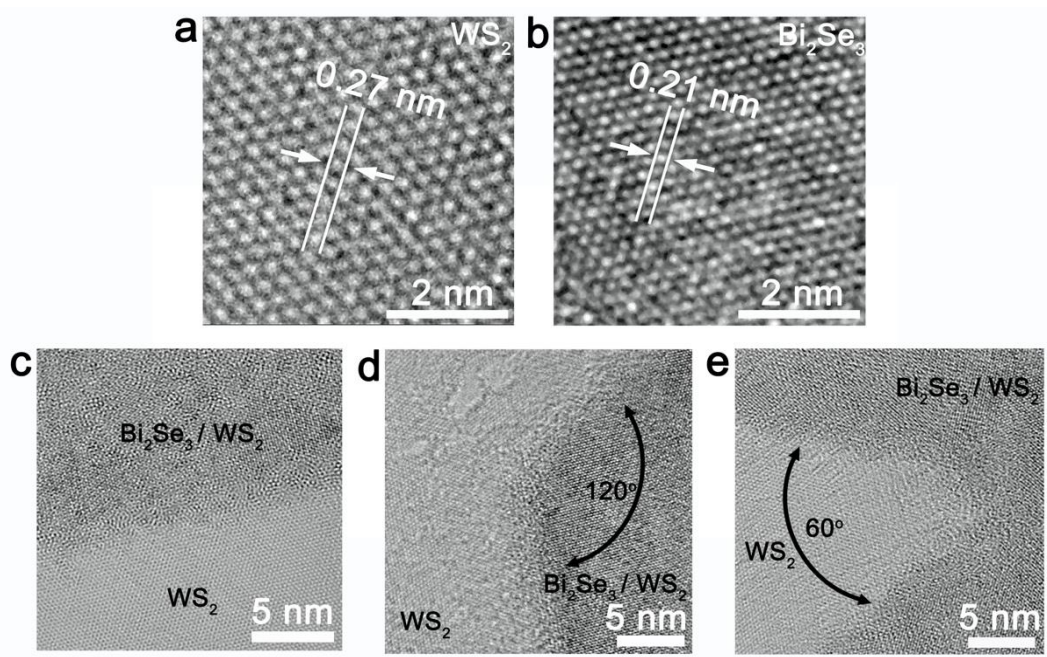

**Figure S12.** The HRTEM images of 2D $Bi_2Se_3/WS_2$ vdW heterostructures. The HRTEM images of $WS_2$ a) and $Bi_2Se_3$ b) layers. c-e) The HRTEM images of the edges of triangular shaped c) and suborbicular shaped d, e) $Bi_2Se_3/WS_2$ heterostructures.

The measured lattice spacings of 0.27 nm ($WS_2$, panel a) and 0.21 nm ($Bi_2Se_3$, panel b) are consistent with the value of $WS_2$ (100) [5] and $Bi_2Se_3$ (110) [6] spacings, respectively. Compared with the atomically flat edges of $Bi_2Se_3$ grown at 600 °C (panel c), the majority of sawtooth edges of $Bi_2Se_3$ grown at 505 °C (panels d and e) possess angles of 60° and 120°, respectively, which reflects the hexagonal lattice and directionality of the chemical bonding.

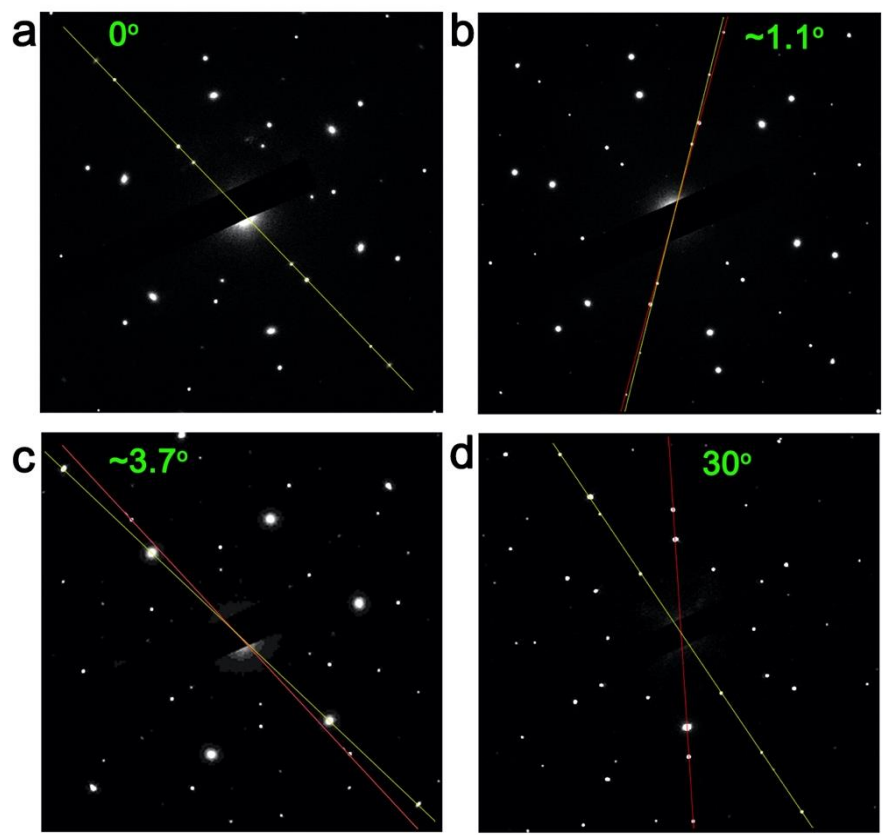

**Figure S13.** SAED patterns of 2D Bi$_2$Se$_3$/WS$_2$ vdW heterostructures with twisting angles of 0°
a), 1.1° b), 3.7° c) and 30° d), respectively.